\documentclass[letter]{aa} % for the letters
\usepackage{amsmath,amstext}
\usepackage{txfonts}
\usepackage{graphicx,subcaption}
\usepackage{color}
\usepackage{multirow}
\usepackage{rotating}
\usepackage{pdflscape}
\usepackage{float}
\usepackage{appendix}
\usepackage{natbib,twoopt}
\usepackage[breaklinks=true]{hyperref} %% to avoid \citeads line fills
\bibpunct{(}{)}{;}{a}{}{,}             %% natbib format for A&A and ApJ
\makeatletter
  \newcommandtwoopt{\citeads}[3][][]{\href{http://adsabs.harvard.edu/abs/#3}%
    {\def\hyper@linkstart##1##2{}%
     \let\hyper@linkend\@empty\citealp[#1][#2]{#3}}}
  \newcommandtwoopt{\citepads}[3][][]{\href{http://adsabs.harvard.edu/abs/#3}%
    {\def\hyper@linkstart##1##2{}%
     \let\hyper@linkend\@empty\citep[#1][#2]{#3}}}
  \newcommandtwoopt{\citetads}[3][][]{\href{http://adsabs.harvard.edu/abs/#3}%
    {\def\hyper@linkstart##1##2{}%
     \let\hyper@linkend\@empty\citet[#1][#2]{#3}}}
  \newcommandtwoopt{\citeyearads}[3][][]%
    {\href{http://adsabs.harvard.edu/abs/#3}
    {\def\hyper@linkstart##1##2{}%
     \let\hyper@linkend\@empty\citeyear[#1][#2]{#3}}}
\makeatother
\newcommand{\Halpha}{\ifmmode {\rm H}\alpha \else H$\alpha$\fi}
\newcommand{\Hbeta}{\ifmmode {\rm H}\beta \else H$\beta$\fi}
\newcommand{\Hgamma}{\ifmmode {\rm H}\gamma \else H$\gamma$\fi}
\newcommand{\Hdelta}{\ifmmode {\rm H}\delta \else H$\delta$\fi}
\newcommand{\Lya}{\ifmmode {\rm Ly}\alpha \else Ly$\alpha$\fi}

\newcommand{\oii}{[O\,\textsc{ii}]}
\newcommand{\oiii}{[O\,\textsc{iii}]}
\newcommand{\zs}{{\it z}}
\newcommand{\rQG}{$r$\textsubscript{0,QG}}
\newcommand{\rQQ}{$r$\textsubscript{0,QQ}}
\newcommand{\rGG}{$r$\textsubscript{0,GG}}
\begin{document}

\title{The web of the Giant: spectroscopic confirmation of a Large Scale Structure around the z=6.31 quasar SDSS~J1030+0524}
%\footnote{Based on XMM-Newton observations}}

   \author{Marco Mignoli\inst{1}
    \thanks{\email{marco.mignoli@inaf.it}}
          \and
          Roberto Gilli\inst{1}
          \and
          Roberto Decarli\inst{1}
          \and
          Eros Vanzella\inst{1}
          \and
          Barbara Balmaverde\inst{2}
          \and
          Nico Cappelluti\inst{3}
          \and\\
          Letizia P. Cassar\`a \inst{4}
          \and
          Andrea Comastri\inst{1}
          \and
          Felice Cusano \inst{1}
          \and
          Kazushi Iwasawa\inst{5,6}
          \and
          Stefano Marchesi\inst{1}
          \and
          Isabella Prandoni\inst{7}
          \and
          Cristian Vignali\inst{8,1}
          \and
          Fabio Vito\inst{9}
          \and
          Giovanni Zamorani\inst{1}
          \and
          Marco Chiaberge\inst{10}
          \and
          Colin Norman\inst{10,11}
   }
   \institute{
   INAF -- Osservatorio di Astrofisica e Scienza delle Spazio di Bologna, OAS, via Gobetti 93/3, I-40129 Bologna, Italy
   \and
   INAF -- Osservatorio Astrofisico di Torino, Via Osservatorio 20, I-10025 Pino Torinese, Italy
   \and
   Department of physics, University of Miami, Coral Gables, FL 33124, USA
   \and
   INAF – Istituto di Astrofisica Spaziale e Fisica Cosmica (IASF), Via A. Corti 12, 20133 Milano, Italy
   \and
   Institut de Ci\`encies del Cosmos (ICCUB), Universitat de Barcelona (IEEC-UB), Mart\'i i Franqu\`es, 1, 08028 Barcelona, Spain
    \and
    ICREA, Pg. Llu\'is Companys 23, 08010 Barcelona, Spain
   \and
   INAF -- Istituto di Radioastronomia, via Gobetti 101, I-40129 Bologna, Italy
   \and
   Dipartimento di Fisica e Astronomia, Universit\`a degli Studi di Bologna, Via Gobetti 93/2, I-40129 Bologna, Italy
   \and
   Scuola Normale Superiore, Piazza dei Cavalieri 7, I-56126 Pisa, Italy
   \and
   Space Telescope Science Institute, 3700 San Martin Dr., Baltimore, MD 21210, USA
   \and
   Johns Hopkins University, 3400 N. Charles Street, Baltimore, MD 21218, USA
   }

   \date{Received  / Accepted  }
%\date{}

\abstract
   {
We report on the spectroscopic confirmation of a large scale structure around the luminous, \zs=6.31 
QSO SDSS~J1030+0524, that is powered by a billion solar mass black hole. The structure is populated
by at least six members, four Lyman Break Galaxies (LBGs) and two Lyman Alpha Emitters (LAEs). The four LBGs 
have been identified among a sample of 21 i-band dropouts with z\textsubscript{AB}$<\,$25.5 selected
up to projected separations of 5 physical Mpc (15 arcmin) from the QSO. Their redshifts have been
determined through up to 8hr-long multi-object spectroscopic observations at 8-10m class telescopes. 
The two LAEs have been identified in a 6hr VLT/MUSE observation centered on the QSO. 
The redshifts of the six galaxies cover the range 6.129-6.355. 
Assuming that peculiar velocities are negligible, this range corresponds to radial separations 
of $\pm$5~physical Mpc from the QSO, that is comparable to the projected scale of the observed LBG
distribution on the sky. We conservatively estimate that this structure is significant 
at $>\,$3.5$\,\sigma$ level, and that the level of the galaxy overdensity is at least 1.5-2 within
the large volume sampled ($\sim$780 physical Mpc$^3$). The spectral properties of the six member galaxies
(\Lya \ strength and UV luminosity) are similar to those of field galaxies at similar redshifts.

This is the first spectroscopic identification of a galaxy overdensity around a super-massive 
black hole in the first billion years of the Universe. 
%While such a single detection is not statistically meaningful, 
Our finding lends support to the idea that the most distant and massive black holes form and 
grow within massive ($>10^{12}\,M_{\odot}$) dark matter halos in large scale structures,
and that the absence of earlier detections of such systems was likely due to observational limitations. 
}

 %galaxies: distances and redshifts --- 

\titlerunning{J1030+0524 LSS}
\authorrunning{Mignoli, Gilli et~al.}

\maketitle

%%%%%%%%%%%%%%%%%%%%%%%%%%%%%%%%%%%%%%%%%%%%%%%%%%%%%%%%%
%%%%%%%%%%%%%%%%%%%%%%%%%%%%%%%%%%%%%%%%%%%%%%%%%%%%%%%%%
%%%%%%%%%%%%%%%%%%%%%%%%%%%%%%%%%%%%%%%%%%%%%%%%%%%%%%%%%

\section{Introduction}\label{sec:intro}

The existence of Super-Massive Black Holes (SMBHs; with masses of 10$^{8-10}$M$_{\odot}$) powering
luminous quasars at \zs$\sim$6 and beyond is a severe challenge for extra-galactic astronomy. 
Theory strongly argues that these objects must have formed and grown within the most massive 
dark matter halos ($M$\textsubscript{halo}$\sim$10$^{12-13}\,M_{\odot}$)
in highly biased regions of the early Universe, where seed black holes may find suitable physical 
conditions to form and sufficiently large reservoirs of gas to grow 
\citepads{2009MNRAS.400..100S, 2018MNRAS.473.4003B, 2019Natur.566...85W}. 
Within these environments, high accretion rates can be triggered and sustained by both frequent mergers
of proto-galaxies and possibly by steady flows of cold gas from which the galaxies can form.
Such early large scale structures (LSSs), whose cores would eventually evolve into local massive
galaxy clusters, are expected to be traced by significant galaxy overdensities that may extend
up to scales of $\sim$10 physical Mpc (pMpc) from the quasar \citepads{2009MNRAS.394..577O}.

The actual number and physical properties of galaxies in early overdensities, however, heavily
depend on the competing processes that promote or prevent galaxy assembly. The same is 
true for their spatial distribution. For instance, both supernovae explosions in these galaxies
and the release of energy from the quasar itself in the form of radiation and gas outflows,
may heat and expel gas from the dark matter halos in the LSS, hampering further star formation
and galaxy assembly \citepads[e.g.][]{2014MNRAS.444.2355C}.
This ``negative feedback" may in fact complicate the detection of these early structures,
especially within a few hundreds kpc from the quasar.
%At present, there is still no direct measurement of the environment around early SMBHs. 
To date, empirical evidence of overdensities around early SMBHs remains elusive.
The attempts to directly measure galaxy overdensities have started since the discovery of high-z quasars,
mainly by selecting Lyman Break Galaxy (LBG) and Lyman Alpha Emitter (LAE) candidates at the QSO redshift, 
but the results have not been conclusive \hbox{\citepads[see e.g.][]{2017ApJ...834...83M}.}
Recently, \citetads{{2018ApJ...856..109O}} reported an example of a candidate LSS
extending for $\sim$4$\times$8~pMpc$^2$ (i.e. 12\arcmin$\times$24\arcmin) in the vicinity of a
\zs=6.6 QSO.  Their Subaru Suprime-Cam narrow and broadband photometric observations select
both LBG and LAE candidates: the low density of LAEs within 3~pMpc from the QSO may hint at
negative feedback acting preferentially on low-mass galaxies hosted by small halos.

Further insight is now coming from ALMA: [CII] observations of a sample of 25 QSOs with 
$M_{BH}\gtrsim 3\times 10^8\, M_{\odot}$ at \zs$>$5.94 revealed that a significant fraction
of them (16\%) are paired to a close (within $\sim$100~kpc) companion galaxy that is rapidly
forming stars \citepads{2017Natur.545..457D}. 
This suggests that early SMBH fueling and galaxy assembly may be favored by galaxy interactions
on very small scales, but the presence of any significant structure on larger scales 
has still to be firmly established \citepads{2019MNRAS.489.1206H}.

With the aim of obtaining the first observational confirmation of an LSS around a super-massive black hole
in the first billion years of the Universe, we started an intensive observational campaign of the field
around the bright QSO SDSS J1030+0524 (z=6.308, $M$\textsubscript{BH}=1.4$\times$10$^9M_{\odot}$; 
\citeads{2007ApJ...669...32K}), where promising evidence for a high-z LSS was reported (see Section 2).

Throughout this Letter we adopt a concordance cosmology with $H_0$ = 70 km s$^{-1}$ Mpc$^{-1}$,
$\Omega_m$= 0.3, and $\Omega_{\Lambda}$ = 0.7, in agreement with the values measured by the
\citetads{2016A&A...594A..13P}. This gives a scale of 5.55 kpc/arcsec at z=6.308. 
All magnitudes are total magnitudes and quoted in the AB system.

%%%%%%%%%%%%%%%%%%%%%%%%%%%%%%%%%%%%%%%%%%%%%%%%%%%%%%%%%
%%%%%%%%%%%%%%%%%%%%%%%%%%%%%%%%%%%%%%%%%%%%%%%%%%%%%%%%%
%%%%%%%%%%%%%%%%%%%%%%%%%%%%%%%%%%%%%%%%%%%%%%%%%%%%%%%%%

\section{SDSS J1030+0524: an over-dense QSO field}\label{sec:field}

\begin{figure}[t]
\centering
\includegraphics[width=9.cm]{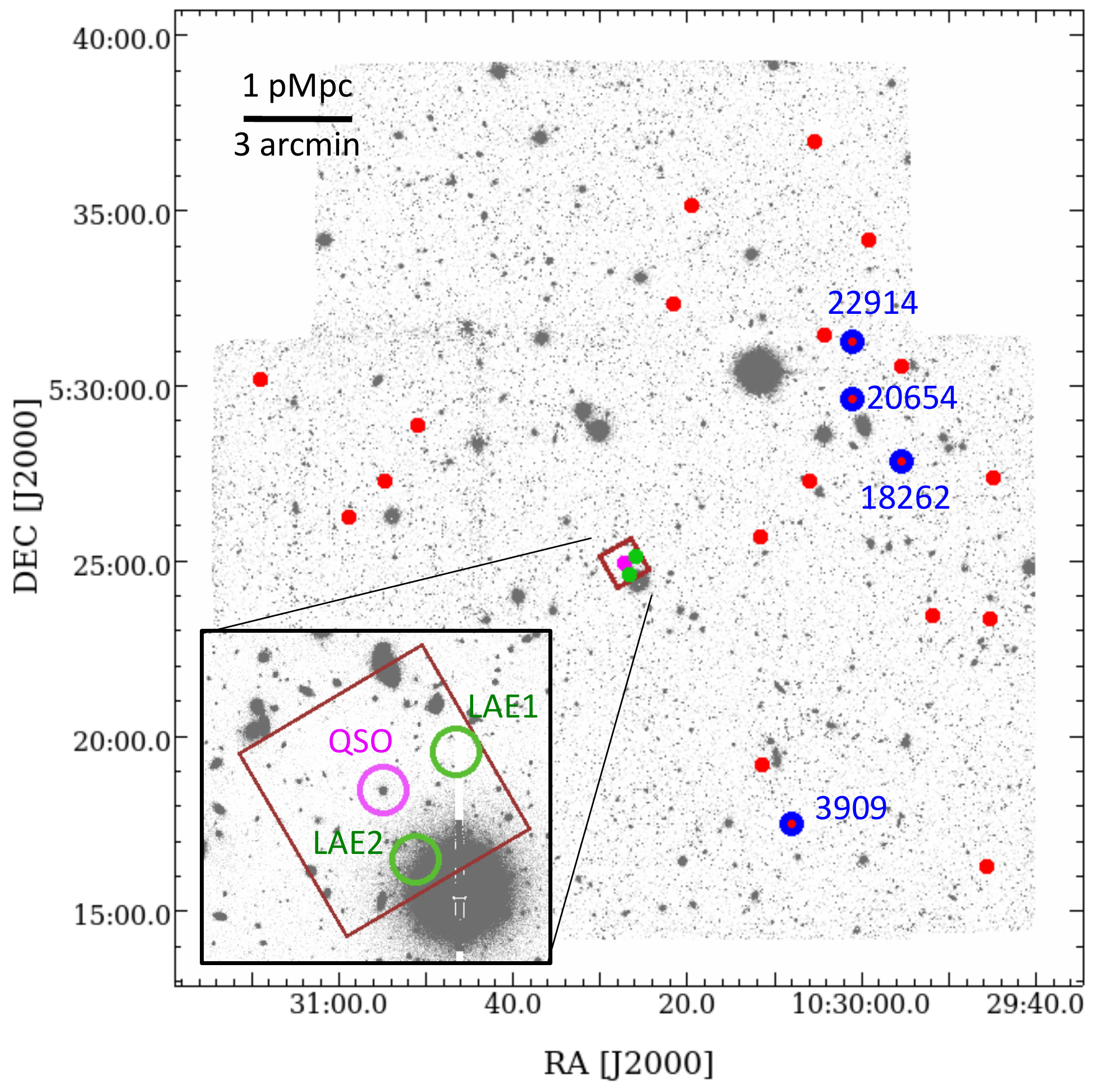}
\caption{Sky distribution of the LBG sample (red dots) overplotted to the LBT/LBC z-band
image of the field. The spectroscopically confirmed members of the LSS at the QSO redshift are marked
in blue and labeled with their ID. The position of the two LAEs in the LSS discovered
by MUSE is marked by green points. The MUSE field of view (FoV) is shown as a brown square. The position of the QSO
is shown in magenta. The inset shows a zoom on the sky region around the MUSE field.}
\label{fig:field}
\end{figure}

Among the $\approx$300~QSOs discovered so far at z$>$5.7 \citepads{2016ApJS..227...11B}, SDSS~J1030+0524 (hereafter J1030)
is hosted in one of the most convincing large-scale galaxy overdensities.
Deep imaging with the 3\arcmin$\times$3\arcmin Advanced Camera for Surveys (ACS) on the Hubble Space
Telescope (HST) revealed a $\sim$3$\sigma$ overdensity of i-band dropouts within $\sim$0.5 projected
pMpc from the QSO \citepads{2005ApJ...622L...1S,2009ApJ...695..809K}. Among all of the \zs$\sim$6
QSO fields observed with HST by \citetads{2009ApJ...695..809K}, J1030 was found to be the most overdense.

In the sample of four luminous \hbox{\zs$\sim$6} QSOs observed with the Large Binocular Camera (LBC)
at the Large Binocular Telescope (LBT) by \citetads{2014A&A...568A...1M}, J1030 was found again to feature
the highest density of i-band dropouts (with density contrast $\delta=\rho/\bar \rho -1 = 2$, at S/N=3.3;
$\rho$ and $\bar \rho$ are the measured and average, background, source density, respectively),
hence reinforcing the result found on (8$\times$) smaller scales by \citetads{2005ApJ...622L...1S}.
In 2016, we obtained deep ($Y_{\rm AB}$=24.5, $J_{\rm AB}$=24) near-IR imaging of the J1030 field 
with the Wide-field InfraRed Camera %(WIRCAM, FoV of 21$\times$21 arcmin$^2$) 
at the Canada France Hawaii 
Telescope\footnote{See http://j1030-field.oas.inaf.it/ for a summary of all data-sets available in the field.}.
Using these new observations, together with archival data from the {\it Spitzer} Infrared Array Camera
(IRAC), \citetads{2017A&A...606A..23B} pushed the selection of \hbox{\zs$\sim$6}
galaxy candidates to fainter fluxes, measured photometric redshifts, and improved the rejection of
contaminants. This analysis allowed the selection of 21 robust \hbox{\zs$\sim$6} candidates down to 
\zs\textsubscript{AB}$\sim$25.7, reinforcing the significance of the LSS overdensity, now with $\delta$=2.4,
at S/N=4.0. Similarly to what observed by HST/ACS on smaller scales, a strong asymmetry
in the spatial distribution of these brighter dropouts was found on larger scales (See Fig.~\ref{fig:field}).

%%%%%%%%%%%%%%%%%%%%%%%%%%%%%%%%%%%%%%%%%%%%%%%%%%%%%%%%%
%%%%%%%%%%%%%%%%%%%%%%%%%%%%%%%%%%%%%%%%%%%%%%%%%%%%%%%%%
%%%%%%%%%%%%%%%%%%%%%%%%%%%%%%%%%%%%%%%%%%%%%%%%%%%%%%%%%

\section{Spectroscopic campaign and UV properties of the galaxy members}\label{sec:data}

We started a systematic program of spectroscopic follow-up of the \hbox{\zs$\sim$6} LBG candidates selected
in \citetads{2017A&A...606A..23B}, to precisely measure their redshifts and to verify if they are actually 
located in the vicinity of the central QSO. 
The campaign used several multi-object spectrographs (MOS):
the DEep Imaging Multi-Object Spectrograph (DEIMOS) on the \hbox{10-m} Keck II telescope
\citepads{2003SPIE.4841.1657F}, the FOcal Reducer and Spectrograph 
\citepads[FORS2,][]{1998Msngr..94....1A} and the Multi Unit
Spectroscopic Explorer \citepads[MUSE,][]{2010SPIE.7735E..08B}, 
both mounted on the Very Large Telescopes (VLTs) at ESO, and, finally, 
the Multi-Object Dual Spectrograph \citepads[MODS,][]{2010SPIE.7735E..0AP}
at the LBT. Details on the observing runs and on the analysis of the spectroscopic
data are presented in Appendix A. 

In summary, we observed 12 out of the 21 LBGs selected by \citetads{2017A&A...606A..23B}:
nine of them resulted in bona-fide high-redshift (\zs>5.7) galaxies, while
the low spectral quality and absence of significant spectral
features prevented any redshift determination for the remaining three. In the archival VLT/MUSE cube 
centered on the QSO, we further detected four LAEs with \zs>5.7: none of them was
selected by \citetads{2017A&A...606A..23B} because they are fainter than the 
magnitude limits of the LBT photometric data. 
We note that the distinction between LBGs and LAEs adopted
for our targets refers essentially to the selection method, since
LBGs may also display Ly$\alpha$ emission. 

\begin{figure}[ht!]
\centering
\includegraphics[ width=8.5cm]{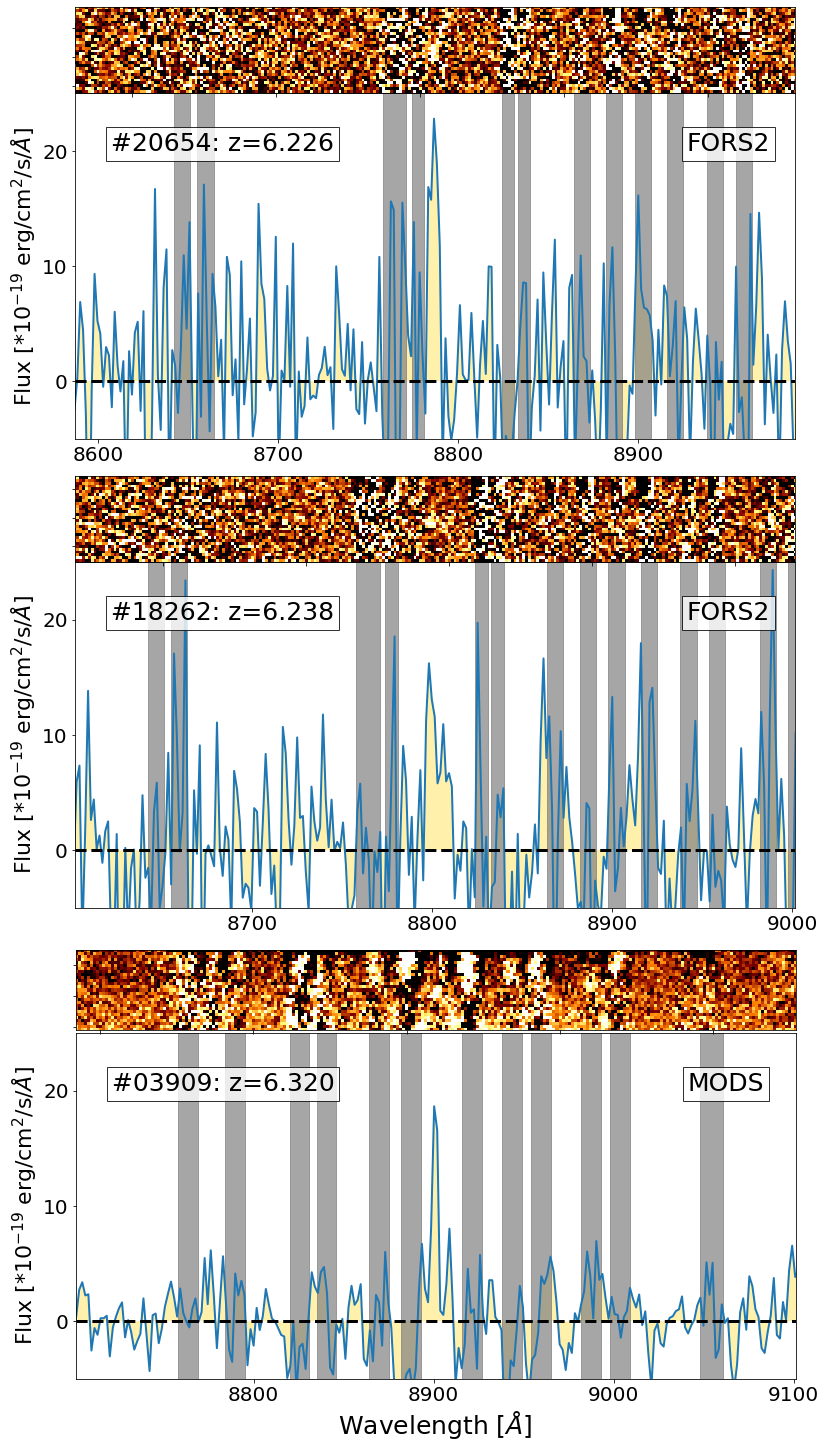}
\caption{Optical spectra of three LBGs in the LSS with redshift measured from the peak of \Lya \ line (see Table~\ref{table:spmeas}).
In each panel the top row shows the 2D-spectrum, while at the bottom the 1D-spectrum
is displayed with gray bands highlighting the spectral regions inaccessible due
to the strong sky line residuals. \label{fig:spec1}}
\end{figure}

\begin{figure}[ht!]
\centering
\includegraphics[ width=8.5cm]{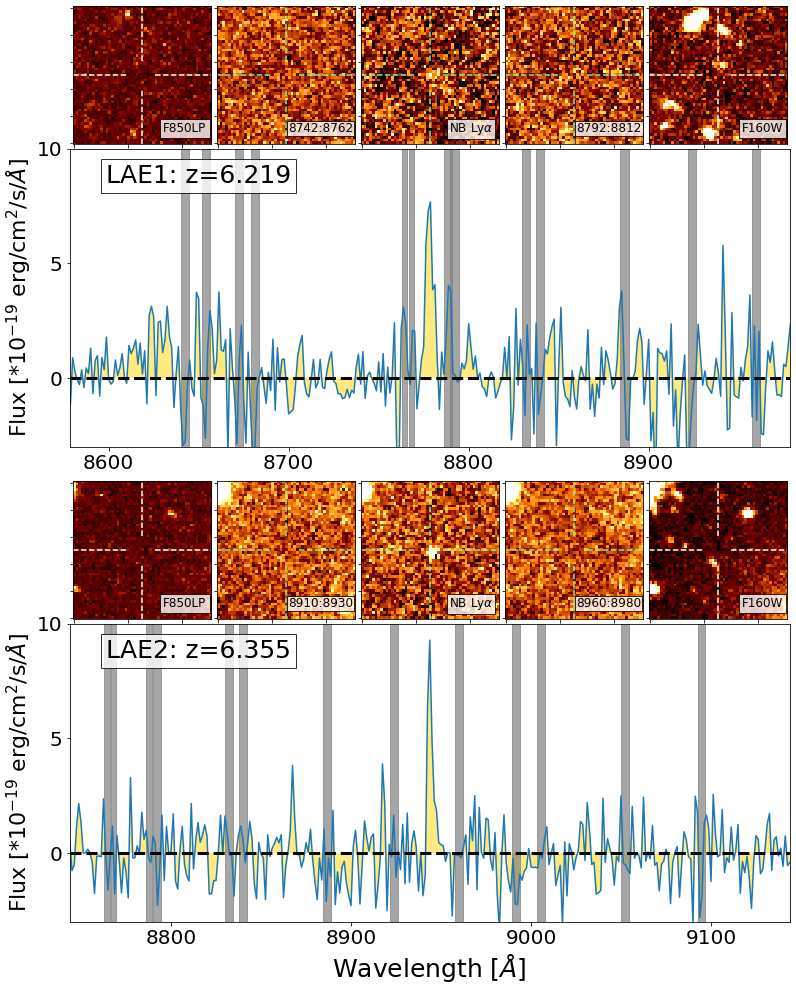}
\caption{MUSE spectra of the two LAEs in the LSS (see Table~\ref{table:spmeas}) are displayed in two panels. 
In each panel the top row shows, from left to right, the HST/ACS F850LP image, three MUSE reconstructed narrow-band 
images (20\AA-wide), and the HST/Wide Field Camera 3 (WFC3) F160W image. The images are 10\arcsec~wide and 
the source position is indicated by cross-hairs. In the bottom row, the extracted 1D-spectrum
is shown in a 400\AA-wide interval around the \Lya \ line, with noisy regions in the MUSE cube affected 
by the sky line subtraction highlighted by gray bands.\label{fig:spec2}}
\end{figure}

\begin{figure}[ht!]
\centering
\includegraphics[width=8.5cm]{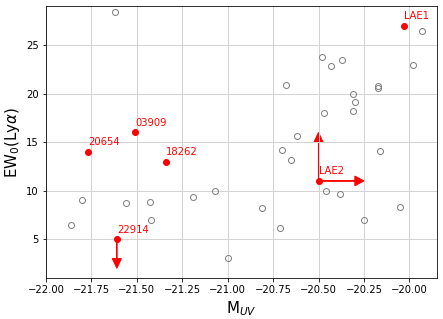}
\caption{EW(\Lya) vs. M\textsubscript{UV} for the six LSS-member galaxies of Table~\ref{table:spmeas}
(red filled circle). For comparison we also plot the same properties from a large sample of
high-z galaxies, presented by \citetads{2017A&A...608A.123D} (gray empty circles).\label{fig:EWvsMUV}}
\end{figure}

We considered the six galaxies with $|z - z_{QSO}|<\,$0.1 (four LBGs and two LAEs, 
see Table~\ref{table:spmeas} and Fig.~\ref{fig:field}) as part of the LSS around the QSO
(see Section~\ref{sec:signi} for the exact procedure adopted to identify LSS members). 
%Neglecting peculiar velocities, this redshift interval corresponds 
%to a maximum radial separation of $\sim$5 pMpc from the QSO, i.e. to the same maximum 
%projected separation of our spectroscopic targets from the QSO (see Fig.~\ref{fig:field}).
The galaxies' spectra are shown in Fig.~\ref{fig:spec1} (for three LBGs; the spectrum of LBG \#22914
has been already published in \citeads{2019A&A...631L..10D}) and Fig.~\ref{fig:spec2} (for the two LAEs).
The spectral and photometric properties of the six galaxies are presented in Table~\ref{table:spmeas}. 
In Fig.~\ref{fig:EWvsMUV} we compare their rest-frame \Lya~equivalent widths (EW) with absolute 
UV magnitudes (M$_{\rm UV}$). Here M$_{\rm UV}$ is 
the monochromatic magnitude at 1350\AA, which, at \hbox{\zs$\approx$6.3},
roughly corresponds to the effective wavelength of the LBC z-band magnitude
from which it was obtained. Although the number of galaxies is limited, we
observe the same trend reported in numerous previous studies of high-z field galaxies:
UV-bright LBGs shows limited \Lya~EWs ($\ll$20\AA), while the fainter LAEs exhibit
larger EWs. Our objects almost overlap with the larger sample of high-z (\hbox{5.4$\,<\,$\zs$\,<\,$6.5})
galaxies collected and analyzed by \citetads{2017A&A...608A.123D}.
A detailed comparison would require further spectroscopic 
identification of galaxies around J1030. Nonetheless, this preliminary analysis 
shows that the UV spectral properties of galaxies in this LSS do not differ significantly 
from those of field galaxies at high redshift.

%%%%%%%%%%%%%%%%%%%%%%%%%%%%%%%%%%%%%%%%%%%%%%%%%%%%%%%%%
%%%%%% Figure of LBGs spectra #20654 #18262 #03909 %%%%%%
%%%%%%%%%%%%%%%%%%%%%%%%%%%%%%%%%%%%%%%%%%%%%%%%%%%%%%%%%

%%%%%%%%%%%%%%%%%%%%%%%%%%%%%%%%%%%%%%%%%%%%%%%%%%%%%%%%%
%%%%%%%%%   Figure of LAE1 & LAE2 MUSE spectra  %%%%%%%%%
%%%%%%%%%%%%%%%%%%%%%%%%%%%%%%%%%%%%%%%%%%%%%%%%%%%%%%%%%

%%%%%%%%%%%%%%%%%%%%%%%%%%%%%%%%%%%%%%%%%%%%%%%%%%%%%%%%%
%%%%%%%%       Table of Identified LBGs    %%%%%%%%%%%%%%
%%%%%%%%%%%%%%%%%%%%%%%%%%%%%%%%%%%%%%%%%%%%%%%%%%%%%%%%%
\begin{table*}
\caption{Identified LBGs \& LAEs}              % title of Table
\label{table:spmeas}        % is used to refer this table in the text
\centering                                % used for centering table
\begin{tabular}{c c c c c c c c}    % centered columns (8 columns)
\hline\hline                        % inserts double horizontal lines
\rule{0pt}{2ex}    
ID  & mag & M\textsubscript{UV} & logSFR & redshift & Instrument & \multicolumn{2}{c}{\Lya} \\    % table heading
    & z-band & & & & & Flux & EW\\
(1) & (2) & (3) & (4) & & & (5) & (6) \\
\hline % inserts single horizontal line   
\noalign{\smallskip}
 22914 & 25.45$\pm$0.22 &-21.61 & 1.70 & 6.319$\pm$0.001 & DEIMOS & $<$3$^{\,a}$ &$<$5 \\
 20654 & 25.27$\pm$0.17 &-21.77 & 1.97 & 6.226$\pm$0.002 & FORS2 & 8.5$\pm$0.2 &14 \\ 
 18262 & 25.70$\pm$0.20 &-21.34 & 1.32 & 6.238$\pm$0.004 & FORS2 & 5.3$\pm$0.3 & 13 \\
 03909 & 25.55$\pm$0.19 &-21.51 & 1.55 & 6.320$\pm$0.003 & MODS & 5.0$\pm$0.3 & 16 \\
 LAE1  & $\approx$27$^{b}$&-20.03 & 0.40 & 6.219$\pm$0.002 & MUSE & 3.7$\pm$0.2 & 27 \\
 LAE2  &$>$26.5$^{c}$&$>$-20.5 & $<$0.6& 6.355$\pm$0.001 & MUSE & 3.0$\pm$0.2 & $>$11 \\
\hline                                             %inserts single line
\end{tabular}
\tablefoot{(1) Source ID;
(2) AB total magnitude. IDs and mags are from \citetads{2017A&A...606A..23B}, except for the 
two MUSE-detected LAEs and ID22914 (see Appendix A); (3) monochromatic magnitude at 1350\AA \ 
(4) logarithm of the Star Formation Rate in units of M$_\odot$/yr, computed from the UV
luminosity adopting the \citetads{1998ARA&A..36..189K} calibration; 
(5) line fluxes in units of 10$^{-18}$ erg/s/cm$^2$; (6) rest-frame
equivalent widths in \AA.\\ $^{a}$~3-$\sigma$ upper limit estimated from the continuum S/N;
$^{b}$~synthetic magnitude estimated by convolving the spectrum
with the LBC z-band filter response; $^{c}$ LAE2 is not detected in the continuum in the
MUSE spectrum and is undetected both in the LBT/LBC and HST photometry, so we adopted the
5$\sigma$ magnitude limit measured in the ACS F850LP image.}
\end{table*} 

%%%%%%%%%%%%%%%%%%%%%%%%%%%%%%%%%%%%%%%%%%%%%%%%%%%%%%%%%
%%%%%%%%%%%%    Figure EW(Lya) vs MUV      %%%%%%%%%%%%%%
%%%%%%%%%%%%%%%%%%%%%%%%%%%%%%%%%%%%%%%%%%%%%%%%%%%%%%%%%

%%%%%%%%%%%%%%%%%%%%%%%%%%%%%%%%%%%%%%%%%%%%%%%%%%%%%%%%%
%%%%%%%%%%%%    Redshift distribution      %%%%%%%%%%%%%%
%%%%%%%%%%%%%%%%%%%%%%%%%%%%%%%%%%%%%%%%%%%%%%%%%%%%%%%%%
\begin{figure}[ht!]
\centering
\includegraphics[width=9.cm]{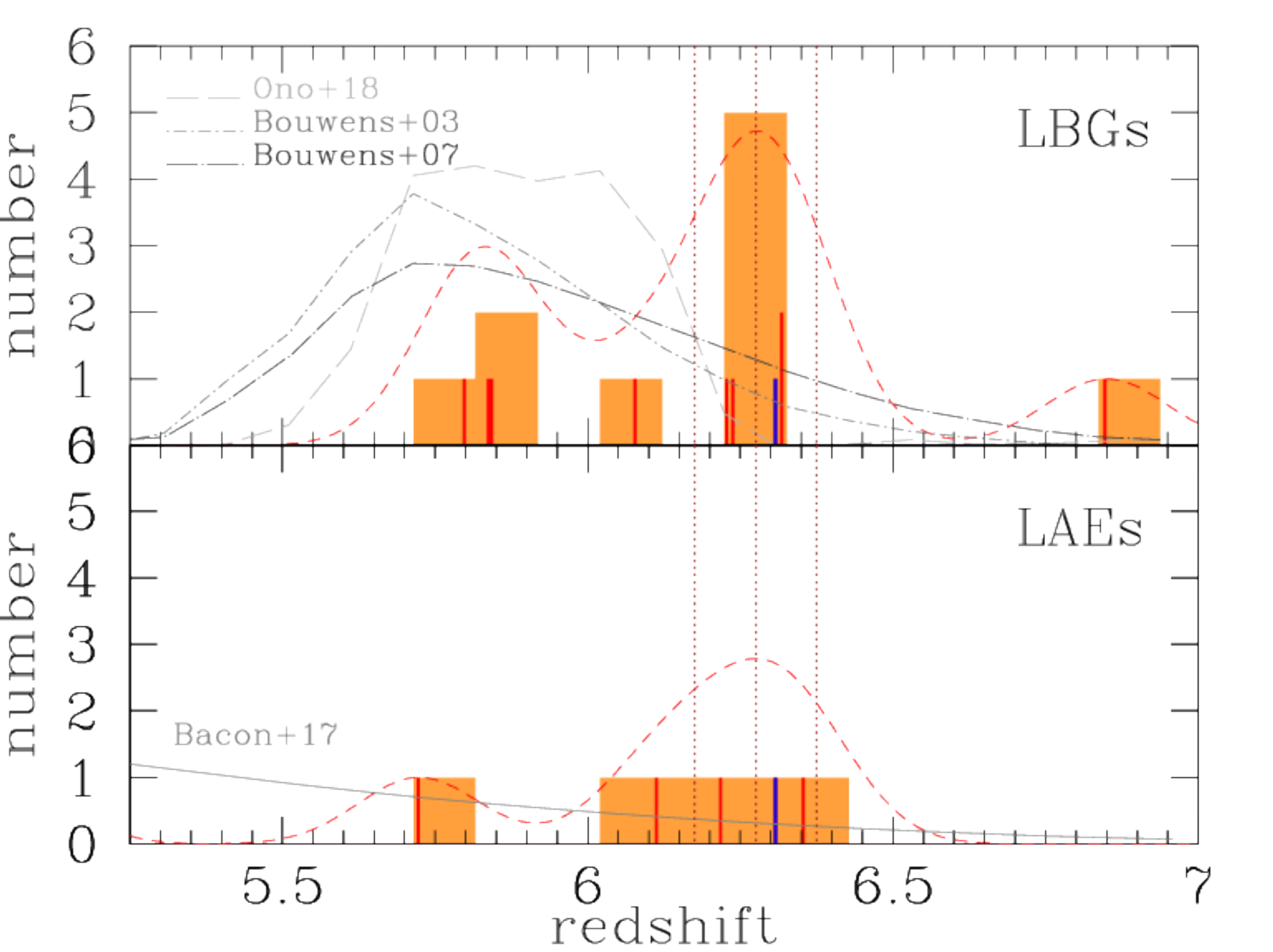}
\caption{{\it Upper panel:} Redshift distribution of all spectroscopically confirmed LBGs 
at \zs$\sim$6 in the J1030 field. The orange and red histogram have bins of 
$\Delta z$=0.1 and 0.005, respectively. The dashed red curve has been obtained by smoothing 
the unbinned distribution with a Gaussian with $\sigma_z$=0.1 (arbitrarily normalized). 
The redshift peak and $\Delta z\pm0.1$ interval used to estimate the structure significance
are shown by the brown vertical dotted lines. The three gray lines show the selection function 
of i-band dropout samples in the literature (see labels) with selection criteria similar to ours. 
{\it Lower Panel:} Same as in the upper panel but for LAEs selected with MUSE. The gray curve
is obtained by heavily smoothing the redshift distribution of LAEs observed with MUSE in the HUDF 
\citepads{2017A&A...608A...1B}. The QSO redshift (marked in blue) is included in the orange
histograms in both panels.}\label{fig:zdist}
\end{figure}

 %%%%%%%%%%%%%%%%%%%%%%%%%%%%%%%%%%%%%%%%%%%%%%%%%%%%%%%%%
 %%%%%%%%%%%%%%%%%%%%%%%%%%%%%%%%%%%%%%%%%%%%%%%%%%%%%%%%%
 %%%%%%%%%%%%%%%%%%%%%%%%%%%%%%%%%%%%%%%%%%%%%%%%%%%%%%%%%
 \section{Redshift distribution and significance of the overdensity}\label{sec:signi}

The redshift distribution of all spectroscopically confirmed LBGs and LAEs at \zs$\sim$6
is shown in Fig.~\ref{fig:zdist} (top and bottom panel, respectively). The distributions are shown in
redshift bins of $\Delta z$=0.1 (orange histogram), as well as in finer, $\Delta z$=0.005, bins (red histogram). 
To isolate potential peaks in the redshift distribution avoiding binning
dependencies, we used a procedure similar to that described in \citetads{2003ApJ...592..721G}. 
We smoothed the unbinned redshift distribution using a Gaussian with $\sigma_z$=0.1 (red dashed line). 
We recall that, at \zs$\sim$6, a separation of 0.1 in redshift space corresponds to a separation of
about 5~pMpc (assuming negligible peculiar velocities). This physical scale corresponds to the maximum 
projected separation between the QSO and the candidate companion galaxies in the LBT/LBC image, 
and is consistent with the observed transverse dimension of other \zs$\sim$6 LSSs
\citepads[e.g.][]{2018ApJ...856..109O}. 
A clear peak at \zs$\simeq$6.275 is observed in both the LBGs and LAEs smoothed distributions 
(red dashed curves in Fig.~\ref{fig:zdist}). We note that the QSO redshift is included in the
determination of the peak position of both redshift distributions (as it would have been selected
in both samples), but not in the computation 
of the structure significance described below. We also note that, for the LBGs, the observed peak 
falls at a significantly higher redshift than that expected by the i-band dropout selection technique
\citepads{2003ApJ...595..589B, 2007ApJ...670..928B, 2018PASJ...70S..10O}, 
suggesting that an LSS is indeed present at \zs$\sim$6.3. 

Once the peak position is determined, we counted all galaxies within $\Delta z\pm$0.1,
i.e. within 5 pMpc, from the peak redshift (brown dotted lines in Fig.~\ref{fig:zdist}). 
The four LBGs and two LAEs of Table~\ref{table:spmeas} fall in this interval.
In order to estimate the expected background of LBGs in that interval we used the selection
function derived in deep HST fields by \citetads{2007ApJ...670..928B}
for i-band dropouts selected by means of color criteria similar to ours
(e.g. {\it i-z}$>$1.3 as the main selection criterion). Two other ``background" curves
are shown in Fig.~\ref{fig:zdist}: one for faint i-band dropouts selected in HST/ACS GTO
fields \citepads{2003ApJ...595..589B} and one for i-band dropouts as bright as ours,
selected as part of the HSC GOLDRUSH project \citepads{2018PASJ...70S..10O}. 
The selection criterion in both \citetads{2003ApJ...595..589B} 
and \citetads{2018PASJ...70S..10O} is {\it i-z}$>$1.5. 
All these background curves have been normalized to 21,
i.e. to the total number of candidate LBGs selected by \citetads{2017A&A...606A..23B}.
Despite the differences produced by the different color cuts adopted, they all show that
the maximum efficiency of i-band dropout selection is between \hbox{\zs$\sim$5.7} and 6. 
By using the \citetads{2007ApJ...670..928B} distribution, we derive a probability of 0.12 that
one galaxy falls in the considered redshift interval. We note that the adopted background
curve has the highest efficiency at \zs=6.3 among the three curves.
%by \citetads{2007ApJ...670..928B}
Therefore, our choice is the most conservative.
Based on the binomial distribution, the probability that 4 (or more) out of 9 spectroscopically confirmed LBGs 
fall in the selected redshift interval, when the expectation of observing one is 0.12, 
is $P_{\rm LBG}$=0.0159. 

We adopted a similar procedure to estimate the binomial probability
that two out of four MUSE LAEs fall in the same redshift interval. To estimate a background source
distribution, we considered the spectroscopic catalog obtained from the 3\arcmin$\times$3\arcmin \ 
MUSE observation of the Hubble Ultra Deep Field \citepads[HUDF,][]{2006AJ....132.1729B}
consisting of 9 pointings of 10~hr each \citepads{2017A&A...608A...1B, 2017A&A...608A...2I}.
We heavily smoothed the MUSE HUDF redshift distribution using a Gaussian with $\sigma_z$=0.5
and then re-normalized it to the average number of galaxies expected in a single pointing
of the HUDF (gray line in the bottom panel of Fig.~\ref{fig:zdist}).
Based on this procedure, we would expect a total of 9 LAEs at \hbox{\zs$>$5.2} in our field, 
which gives a probability of observing one in the considered redshift interval equal to 0.062.
We note that if anything, this procedure probably overestimates the average galaxy background
density because the HUDF observations are deeper than those in the J1030 field and then likely
return a redshift distribution skewed towards slightly higher redshifts. Hence, the derived
significance may again be regarded as conservative. 
Based on the binomial distribution, the probability that 2 (or more) out of 4 spectroscopically confirmed LAEs 
fall in the considered redshift interval, when the probability of observing one is 0.062,
is $P_{\rm LAE}$=0.0212. 

By simply combining the two probabilities computed above, we finally
estimate that the structure is detected at the $1-P_{\rm LBG}\times P_{\rm LAE}=0.9997$ confidence level, 
which corresponds to 3.5$\sigma$ in Gaussian statistics.
We double checked these numbers by running Monte Carlo simulations drawing 9 LBGs and 4 LAEs from the 
redshift probability distributions described above. We run $10^5$ realizations and checked how many times 
four or more LBGs and two or more LAEs fall in the considered redshift interval. The probability values 
obtained by the simulations are identical to those obtained with the binomial statistics.

 %%%%%%%%%%%%%%%%%%%%%%%%%%%%%%%%%%%%%%%%%%%%%%%%%%%%%%%%%
 %%%%%%%%%%%%%%%%%%%%%%%%%%%%%%%%%%%%%%%%%%%%%%%%%%%%%%%%%
 %%%%%%%%%%%%%%%%%%%%%%%%%%%%%%%%%%%%%%%%%%%%%%%%%%%%%%%%%
\section{Discussion and Conclusions}\label{sec:discu}

\subsection {Robustness of the structure detection} 

We show here that the 3.5$\sigma$ level derived in the previous section is a robust lower limit to the LSS detection significance.
 
First we note that there is no bias in our observations towards sources at \zs$\sim$6.3.
From Fig.~\ref{fig:spec1} it is evident that a significant fraction of the red
%($\gtrsim$8500\AA)
spectral range is blinded by the noise produced by the subtraction of the strong night-sky lines,
making the detection of faint \Lya \ lines highly inefficient. The distribution of sky lines is
nonetheless rather uniform across the \zs=5-7 redshift range. As the probability of successfully
measuring the redshift of a source inside or outside the structure is the same, we then observe
a number of redshifts in and out the structure that have been likely depressed by the same factor. 
Accounting for this effect would increase the statistical significance of the LSS detection.
Also, we did not fine tune the smoothing length and width of the structure interval to maximize
the signal. As an example, repeating the computations using $\delta$\textsubscript{\zs}$\sim\,$0.08
around the peak redshift, would increase the significance to 3.7$\sigma$. 
In our significance computations we conservatively never included the QSO. 
However, the QSO is part of the structure and it would have been selected in both the LBG and LAE
samples discussed above. If we include it in the LBG sample and compute the binomial
probability that 5 out of 10 sources extracted from the same probability distribution adopted above
%\citepads[from][]{2007ApJ...670..928B} 
fall in the considered redshift range, we obtain $P_{\rm LBG}$=0.0037. 
When combined with $P_{\rm LAE}$ computed above this gives a joint significance of 3.8$\sigma$. 
Finally, in our computations we considered among the possible reference background curves, those that 
are more skewed towards the highest redshifts, i.e. those returning the highest background probability 
of observing a galaxy around \zs$\sim$6.3. This choice is very conservative. For instance, by adopting 
the redshift distribution of i-band dropouts of \citetads{2018PASJ...70S..10O}, which shows a sharp
drop beyond z=6, we would estimate a joint significance of $>4\sigma$. 

\subsection{Overdensity level}

At present, the number of LBGs that have been spectroscopically identified as part of the LSS 
around the QSO is $n=4$, whereas one would expect an average of $\bar n \sim 2$ assuming the 
\citetads{2007ApJ...670..928B} redshift probability distribution normalized to 21, i.e. the total 
number of LBGs in the J1030 field \citepads{2017A&A...606A..23B}. This corresponds to a lower limit to 
the measured level of the spectroscopic overdensity of \hbox{$\delta = n/ \bar n - 1 \geq\,$1}  
($\geq$1.5 if the QSO is added), as subsequent successful identifications of LBGs at
\hbox{\zs$\sim$6.3} from the \citetads{2017A&A...606A..23B} sample will simply increase $n$. 
Also, the total number of LBGs in the field is likely enhanced by the presence of the LSS and
the adopted value for $\bar n$ presumably overestimates the true average density.
Consequently, the lower limit to $\delta$ estimated above is conservative (based on similar arguments,
we derive a lower limit to the LAE overdensity of $\delta$\textsubscript{LAE}$\,\geq\,$2.3). 
We note that the estimated overdensity level is measured over a remarkably large volume of $\sim$780 pMpc$^3$,
and is consistent, within the statistical errors, with what was measured through imaging by \citetads{2017A&A...606A..23B}. 
 
Based on the observed overdensity, we provide a rough estimate of the clustering level of early galaxies 
and QSOs that is required to explain it. To this aim, the expected overdensity level can be obtained 
by integrating the LBG-QSO cross-correlation function (assumed here to be a power-law 
$\xi(r)=(r/\rQG)^{-\gamma}$ with cross-correlation length \rQG) over the considered volume. 
For simplicity, we use for this computation a sphere of radius $r_s$=5.7 pMpc around the QSO, 
which contains all the LBGs in the LSS, and has the same volume considered in the previous sections 
(780 pMpc$^3$, i.e. a cylinder of both radius and half-length 5 pMpc). 
This gives $\delta(<r_s) = 3\xi(r_s)/(3-\gamma)$.  
An overdensity of $\delta>$1.0$-$1.5, as observed e.g. for our LBGs, would require
\rQG$\,>\,$24$-$29~Mpc comoving (cMpc) for $\gamma$=2.  Assuming that the
auto-correlation functions of QSOs and galaxies are both power-laws with the same slope, the relation 
\rQG=(\rQQ$\times$\rGG)$^{1/2}$ holds, where \rQQ \ and \rGG \ are
the auto-correlation lengths of QSOs and galaxies, respectively. As \rGG$\approx$ 20$-$30~cMpc 
for galaxies at \zs$\sim6$ with $M_{UV}$ and star formation rate (SFR) similar to ours \citepads{2019MNRAS.489..555K}, 
we estimate that \rQQ \ is at least $\sim$20~cMpc. Such a large correlation length at \zs$\sim$6 
in turn suggests that J1030 is hosted by a dark matter halo with mass 
$M$\textsubscript{h}$\,>\,$10$^{12}\;M_{\odot}$ \citepads{2018PASJ...70S..11H, 2019MNRAS.489..555K}. 
 
\subsection{Environment of early black holes: future prospects}

Although simulations suggest that SMBHs in the early Universe should reside on average in the most 
massive ($10^{12-13}\,M_{\odot}$) dark matter halos formed at that time, a large variance in the 
galaxy number counts around them is nonetheless expected, and some SMBHs may not even show enhanced
counts in their neighborhoods \citepads{2019MNRAS.489.1206H}. 
The expected variance is even larger when feedback effects are considered, 
either from the QSO itself or from strong stellar winds in the companion galaxies 
\citepads{2014MNRAS.444.2355C}. 
Such feedback effects may suppress star formation, and hence decrease the luminosity and stellar mass 
of the galaxy members within scales of 1-2 pMpc, which would then fall below the sensitivity
of most observations of high-z QSO fields. 

To obtain a reasonable census of the environment of high-z QSOs, 
explorations of deep-and-wide areas around several such systems is needed. Our approach based 
on LBT/LBC observations is along these lines 
\citepads[see also][for deep-and-wide observations with Subaru/Suprime-Cam]{2018ApJ...856..109O}.
Expensive follow-up observations with optical spectrographs at 8-10m class telescopes are needed to 
secure redshifts of the relatively luminous LBG candidates selected over wide areas, as these may
have weak \Lya \ lines. As shown in Fig.~\ref{fig:spec1}, such observations are complicated
by the presence of strong sky lines. High-z LAEs, on the other hand, can be nowadays efficiently selected 
with MUSE down to very faint fluxes. The main limitation with such an approach is the small FoV of MUSE, 
which requires a large number of pointings.  
The analysis presented in this paper shows that obtaining a robust estimate of the galaxy density
around early QSOs is an extremely time-consuming process, even for a single system.

A complementary approach to obtain accurate redshifts is searching for the [CII]158$\,\mu$m
or the  [OIII]88$\,\mu$m line in spectroscopic observations at (sub-)mm frequencies 
\citepads[e.g.][]{2017Natur.545..457D, 2019PASJ...71..109H, 2020ApJ...896...93H}. 
This approach is particularly promising as it allows for a redshift estimate in those LBG 
candidates with absent or weak \Lya \ \citepads{2019A&A...631L..10D}, or falling in redshift
windows covered by sky lines (see Fig.\ref{fig:spec1}). Also, it allows for the discovery of dusty
companion galaxies that cannot be selected at any other wavelength \citepads{2017Natur.545..457D}. 
ALMA observations are then an extremely promising tool to identify members of LSSs around 
\zs$\sim$6 QSOs.
In particular, for the J1030 field, such observations may increase the statistical
significance of the overdensity reported in this paper and, importantly, deliver a sizeable sample
of confirmed members that we can use to identify common trends and/or systematic differences
in the physical properties of LSS and field galaxies at \zs$\sim$6.

\begin{acknowledgements}
We thank the referee for careful reading and detailed report that improved the quality of the manuscript.
We acknowledge funding from the INAF main-stream (1.05.01.86.31) and from the agreement ASI-INAF n. 2017-14-H.O.
This work is based on ESO program \hbox{0102.A-0263(A)}, on observations obtained at the W.~M.~Keck Observatory,
%which is operated jointly by the California Institute of Technology and the University of California
and on data acquired using the Large Binocular Telescope (ID 2017/2018 \#18).
\end{acknowledgements}

\bibliographystyle{aa}

\begin{appendix}
\section{Observing runs and spectral analysis}

%%%%%%%%%%%%%%%%%%%%%%%%%%%%%%%%%%%%%%%%%%%%%%%%%%%%%%%%%
%%%%%%%%%%%%     Table of Observations      %%%%%%%%%%%%%
%%%%%%%%%%%%%%%%%%%%%%%%%%%%%%%%%%%%%%%%%%%%%%%%%%%%%%%%%
\begin{table}[b]
\setlength{\tabcolsep}{3pt}
\caption{Table of spectroscopic observations             % title of Table
\label{table:obslog}}                % is used to refer this table in the text
\centering
\begin{footnotesize}
\begin{tabular}{ccccc}
\hline \hline
\rule{0pt}{2ex}    
 Telescope & Instrument & Date & Exp.Time & Spectral Range\\
 & & & & (1) \\
\hline
\rule{0pt}{2.5ex}    
VLT & MUSE & 2015-2016 & 23040s & 4750$-$9350$\,$\AA\\
Keck & DEIMOS & Feb 2017 & 14400s & $\approx$7000$-$10500$\,$\AA \\
VLT & FORS2 & Jan-Mar 2019 & 29055s & $\approx$7000$-$11000$\,$\AA \\
LBT & MODS & 2018-2019 & 14400s & $\approx$5400$-$10000$\,$\AA \\
\end{tabular}
\tablefoot{(1) The spectral ranges reported for the slit spectrographs are indicative,
as they depend on the geometrical position of the slit within the
mask used for the MOS observation.}
\end{footnotesize}
\end{table}

\subsection{DEIMOS observations}

We observed our \hbox{\zs$\sim$6} candidates with DEIMOS on Feb 27, 2017.
We used the 830 lines~mm$^{-1}$ grating, and the OG550 order cut filter with
the central wavelength of 8500\AA, to efficiently cover the red spectral range
$>$7000\AA. Because of bad weather, we observed only one of the two designed masks,
and the exposure time was reduced to four hours instead of the planned six. 
Nine candidate high-z galaxies from \citetads{2017A&A...606A..23B} were included
in the observed slitmask: three of them were confirmed to be genuine
high-redshift (\zs$>$5.7) LBGs, other two showed very faint (S/N$\lesssim$2)
features compatible with a redshift larger than six (but one of the two unsecure 
identifications was LBG \#20654, subsequently confirmed by deeper FORS2 observations),
while the remaining four were too noisy or even undetected. We also included in the masks
an LBG candidate, \#22914, that was not part of the original sample of \citetads{2017A&A...606A..23B}, 
but was subsequently included after revision of the photometric errors.
We measured for LBG \#22914 a redshift of \zs=6.319 based on a clear Lyman break in the spectrum.
The redshift was later confirmed by the detection of a [CII]158$\,\mu$m line in the NOEMA mm-band
spectrum \citepads[][the DEIMOS spectrum is shown in their Fig.~1]{2019A&A...631L..10D}. 

\subsection{FORS2 observations}
 
During the ESO observing period P102 we were granted VLT/FORS2 observations
to obtain spectra of the \zs$\sim$6 LBG candidates with three 10 hours-long
masks. In the first months of 2019 around one third of the project was
completed in service mode. We used FORS2 equipped with MIT red-optimized CCD and
adopted grism 600Z+23 to observe the wavelength range 7000$-$11000\AA \
covering the redshifted \Lya \ position for \zs$>$5.15.
The only and partially (80\%) completed mask included 8 primary targets: four of them
have been spectroscopically confirmed as high-z LBGs and two of them reside
in the \hbox{\zs$\simeq$6.3} structure. The two upper panels of Fig.~\ref{fig:spec1}
show portions of the 2-D and 1-D spectra covering the only spectral feature 
visible in the full observed spectral range: the well detected 
emission line is identified as \Lya. We can confidently rule out other possible line
identifications and redshift solutions. The low-z solutions, with 
the emission line associated with either \Halpha, \oiii$\lambda$5007, or \Hbeta,
are discarded due to the lack of the expected nearby lines and because the broadband
colors of these solutions %star-forming galaxies with relatively blue continua
are generally inconsistent with the $i$-dropout selection. 
The most plausible 
alternative identification is \oii$\lambda$3727 at \hbox{\zs$\simeq$1.4},
but this emission line would be resolved into a doublet at our spectral resolution, 
which is not observed. In the LBG~\#20654 spectrum an unresolved
emission line is detected with a formal S/N=9 at the observed wavelength of 8787\AA,
placing this objects at \zs$=$6.226. The line is also detected, albeit
with much lower significance, in a DEIMOS spectrum, further confirming its authenticity. 
The FORS2 spectrum of LBG~\#18262 shows a slightly resolved
emission line, detected with S/N=8 at the observed wavelength of 8801\AA. 
The line is undoubtedly asymmetric with a red wing, a typical shape of
high-redshift \Lya \ emission lines, and yields a redshift of 
\zs$=$6.238 for this source.

\subsection{MODS observations}

In 2017 we were granted an INAF-LBT Strategic Program\footnote{ID 2017/2018 \#18 (P.I. R. Gilli).} 
to identify the X-ray sources detected with a
$\sim$500~ks Chandra exposure in the J1030 field \citepads{2020A&A...637A..52N}.
As part of this large (52 hours) optical/NIR program, we observed in 2018 and 2019 a total of
9 MODS masks mainly dedicated to spectroscopically follow up X-ray sources. 
The observations were obtained in dichroic mode to obtain blue (with the G400L grating)
and red (with G670L grating) spectra simultaneously on the blue and red 
spectrograph channels. Since the density
of the X-ray counterparts was well matched with the number of slits that can be placed
in each mask, only few \hbox{\zs$\sim$6} LBG candidates were included as fillers
in this program. Therefore, and also because of the limited exposure time per mask (4hr)
only one LBG was identified through MODS observations.
The MODS spectrum of LBG~\#03909 (bottom panel of Fig.\ref{fig:spec1}) 
shows an unresolved emission line detected with S/N=6 at the observed wavelength 
of 8901\AA, placing this objects at \hbox{\zs$=$6.320.} 
The same considerations given above for the FORS2 observations exclude emission lines
other than \Lya.  

\subsection{Archival MUSE observations}
The J1030 field was observed with MUSE in April 2015 and then again in January 2016
under the ESO program ID~095.A-0714 (PI Karman). MUSE is an integral field spectrograph
with a one~arcmin$^2$ field-of-view and a spatial sampling of 0.2$\times$0.2~arcsec$^2$,
covering the wavelength range 4750$-$9350{\,}\AA \ with a spectral bin of 
\hbox{1.25{\,}\AA{\,}pixel$^{-1}$.} The data reduction and analysis of the archival
6.4hr MUSE data are presented in \citetads{2019A&A...632A..26G}. 
In brief, we measured a redshift for 102 objects (16 of them at \hbox{\zs$>$4)}
in the square arcmin region around the central QSO, a region that did not include 
any of the LBG candidates selected by \citetads{2017A&A...606A..23B}. In spite of that,
MUSE observations revealed four Lyman Alpha Emitters (LAEs) at \zs$>$5.7, 
with two of them at \hbox{\zs$\approx$6.3}.
The MUSE observations of LAE1 is shown in the top panel of Fig.\ref{fig:spec2}. 
In the 1-D spectrum an unresolved emission line is detected with S/N=9 at the
observed wavelength of 8778\AA, placing this objects at \hbox{\zs$=$6.219.} 
The MUSE spectrum of LAE2 shows a strong
unresolved emission line detected with S/N=10 at the observed wavelength 
of 8944\AA. The line is asymmetric with a red wing and yields a redshift of 
\hbox{\zs$=$6.355}. The source is not detected in the continuum, nor in
the HST F850LP and F160W filters. 
The non-detections of MUSE-selected LAEs in medium-deep HST images
is not uncommon, as shown by the MUSE-Wide Survey, where 55\% of \hbox{\zs$>$2.9}
LAEs are undetected in deep CANDELS photometric catalogs \citepads{2019A&A...624A.141U}.

\subsection{Spectra calibration and analysis}
The spectra of the LBGs targeted with MOS instruments were flux-calibrated
using observations of spectro-photometric standard stars obtained during the observing runs.
The spectra were further calibrated  using the z-band magnitudes presented in
Table~\ref{table:spmeas}, by integrating the spectra over the known LBC filter bandpasses.
Since all the observations were performed in good seeing conditions ($\lesssim$1\arcsec)
and with slit-width of 1$-$1.2\arcsec, the relatively small corrections amount up to $\approx$50\%.
The absolute flux calibration of the MUSE data is very good \citepads{2017ApJ...834..174K}
and the absence of flux losses is further demonstrated by the excellent agreement between
the optical magnitudes obtained from our LBC broad-band photometry and the
synthetic magnitudes estimated from MUSE spectra.
The redshift of all the galaxies presented for the first time in this paper has been measured
from position of the \Lya \ emission peak. The redshift value is consistent, within the errors,
with that obtained from the Gaussian fit of the line, apart from the two objects (LBG~\#18262 and LAE2)
in which \Lya \ has an asymmetrical profile and the peak position is known to provide a more correct
redshift estimation.
The fluxes of the \Lya \ emission of Table~\ref{table:spmeas} were measured by
integrating the flux over the line profile.	
The upper limits of flux and equivalent width of the \Lya \ line in \#22914
were estimated from the continuum S/N and the spectral resolution following 
the recipe of \citetads{2009A&A...493...39M}.
LAE1 is not detected in the LBT/LBC nor in the HST photometry, but a very low
S/N continuum redward of the emission line is marginally detected in the MUSE spectrum,
from which we derived a synthetic magnitude {\it z}\textsubscript{AB} of $\approx$27 
by integrating the continuum convolved with the LBC z-band filter response.
This value is compatible with the non-detection of the source in the HST/ACS F850LP image 
%(leftmost box in the upper row; AB$_{5\sigma}\sim\,$26.5).
(leftmost box in the upper row of the LAE1 panel in Fig.~\ref{fig:spec2}; AB$_{5\sigma}\sim\,$26.5).
LAE2 is not detected in the continuum neither in the MUSE spectrum nor in the
HST photometry, then we used the 5-$\sigma$ magnitude limit of 26.5
estimated from the HST/ACS F850LP image to constrain the UV continuum flux.
This upper limit in the continuum yields the lower limit of 
the \Lya~equivalent~width quoted in Table~\ref{table:spmeas}. 

\end{appendix}

\end{document}